\documentclass{aastex}
\usepackage{spr-astr-addons}
\usepackage{url}

\RequirePackage{color}

\begin{document}

\title{A class of solutions for anisotropic stars admitting conformal motion}

\shorttitle{A class of solutions for anisotropic stars}
\shortauthors{Rahaman et al.}

\author{Farook Rahaman\altaffilmark{1}}
\and
\author{Mubasher Jamil\altaffilmark{2}}
\and
\author{Ranjan Sharma\altaffilmark{3}}
\and
\author{Kausik Chakraborty\altaffilmark{4}}

\altaffiltext{1}{Department of Mathematics, Jadavpur University, Kolkata, India. Email: farook$\_$rahaman@yahoo.com}
\altaffiltext{2}{Center for Advanced Mathematics and Physics, National University of Sciences and Technology, Rawalpindi, 46000, Pakistan. Email: mjamil@camp.nust.edu.pk}
\altaffiltext{3}{Department of Physics, St. Joseph's College,
Darjeeling-734 430, India. Email: rsharma@iucaa.ernet.in}
\altaffiltext{4}{Department of Physics, Government Training College, Hooghly - 712103, West Bengal, India. Email: kchakraborty28@yahoo.com}

\begin{abstract}
We provide a new class of interior solutions for anisotropic  stars
admitting conformal motion. The Einstein's field equations in this
construction are solved for specific choices of the density/mass
functions. We analyze the behavior of  the model parameters like
radial and transverse pressures, density and surface tension.
\end{abstract}


\section{Introduction}
In the modelling of compact objects it is generally assumed  that
the underlying matter distribution is homogeneous and isotropic
i.e., a perfect fluid \citep{clayton,kipp}. Such an approach is, in
general, adopted to model polytropic stars like white dwarfs,
compact objects like neutron stars and ultra-compact objects like
strange stars \citep{glen}. However, theoretical advances show that
pressure inside a compact object need not be completely isotropic
and various factors may contribute to pressure anisotropy
\citep{ruder,can,dev1,dev2}. Consequently, the pressure inside a
compact star may be decomposed into two components: the radial
pressure $(p_r)$ and the transverse pressure $(p_t)$. The later acts
in the orthogonal direction to the former one and their difference
$\Delta=(p_t-p_r)$, is crucial in the calculations of surface
tension of a  compact star \citep{Sharma2}. \citet{ruder} showed
that nuclear matter may become anisotropic in the high density
region of order $10^{15}$ gm/cc, which is expected at the core of
compact terrestrial objects. Though we lack a complete understanding
of the microscopic origin of the pressure anisotropy, the role of
pressure anisotropy in the modeling of compact stars is a field of
active research. In particular, it has been predicted that
anisotropy plays a crucial role in the modeling of ultra-compact
stars like strange stars \citep{Aktas}. \citet{witten} suggested
that if quarks are de-confined from the hadrons then $u$, $d$ and $s$
quarks may yield a stable ground state of matter. This state of
matter is termed as strange matter and a star composed of strange
matter is called a strange star. Recent observational and empirical
findings related to several compact objects like Her X-1, SAX J
1808.4-3658, RX J185635-3754 and PSR 0943+10 strongly suggest that
they are strange stars \citep{bombaci,xu}. There are several ways a
strange star can form: a massive star may go under core collapse
after a supernova explosion; alternatively a rapidly spinning star
may  undergo a phase transition to become a strange star. Since the
density inside a strange star is beyond nuclear density anisotropy
may develop and, therefore, a relativistic treatment with
anisotropic pressure should be a reasonable approach to model such
stars.

In the present work, we provide new solutions to model anisotropic
stars admitting conformal motion. The plan of the paper is as
follows. In section $2$ we provide the basic equations which
describe an  anisotropic stars admitting conformal motion. In
sub-sections $2.1$ and $2.2$, we solve these equations for a
specific choices of the density/mass functions. We conclude by
discussing our results in section $3$.

\section{Anisotropic stars}
Inspired by some earlier works \citep{mak,Aktas,Rahaman} on
anisotropic stars admitting a one parameter group of conformal
motions, we look for a new class of anisotropic star solutions
admitting conformal motion. It is well known that to find the
natural relation between geometry and matter through the Einstein's
equations, it is useful to use the inheritance symmetry. The well
known inheritance symmetry is the symmetry under conformal killing
vectors(CKV) i.e.,
\begin{equation}
L_\xi g_{\mu\nu} = \psi g_{\mu\nu},\ \ \ \mu,\nu=1,2,3,4.\label{eq1}
\end{equation}
The quantity on the left hand side is the Lie derivative of the
metric tensor, describing the interior gravitational field of a
compact star with respect to the vector field $\xi$, and $\psi(r)$
is an arbitrary function of $r$. If $\psi$ is a constant then
equation~(\ref{eq1}) generates homoetheties while $\psi=0$ results
in killing vectors. Conformal killing vectors provide a deeper
insight into the spacetime geometry.

To generate a new class of solutions making use of this symmetry,
we choose the static spherically symmetric spacetime in the standard
form (chosen units are $c=1=G$)
\begin{equation}
ds^2 =   e^{\nu(r)} dt^2-e^{\lambda(r)} dr^2-r^2( d\theta^2+
\sin^2\theta d\phi^2),\label{eq2}
\end{equation}
where $\nu(r)$ and $\lambda(r)$ are yet to be determined.

For an anisotropic matter distribution, with the energy-momentum
tensor given by $T_{ij}= \mbox{diag}~(\rho,~-p_{r},~
-p_{t},~-p_{t})$, the Einstein's field equations for the metric
(\ref{eq2}) are obtained as
\begin{eqnarray}
e^{-\lambda}\left[\frac{\lambda^\prime}{r} - \frac{1}{r^2}\right]
+\frac{1}{r^2}&=& 8\pi \rho, \label{eq3}\\
e^{-\lambda}\left[\frac{1}{r^2}+\frac{\nu^\prime}{r}\right]-
\frac{1}{r^2}&=& 8\pi p_r, \label{eq4}\\
\frac{1}{2} e^{-\lambda}\left[\frac{1}{2}(\nu^\prime)^2+
\nu^{\prime\prime}-\frac{1}{2}\lambda^\prime\nu^\prime +
\frac{1}{r}({\nu^\prime-\lambda^\prime})\right]& =& 8\pi p_t,\label{eq5}
\end{eqnarray}
where a $\prime$ denotes differentiation with respect to $r$. Now, the equation
\begin{equation}
L_\xi g_{\mu\nu} =\xi_{\mu;\nu}+ \xi_{\nu;\mu} = \psi g_{\mu\nu}, \label{eq6}
\end{equation}
for the line element given in equation~(\ref{eq2}) generates
\begin{eqnarray}
\xi^1 \nu^\prime &=& \psi, \label{eq7}\\
\xi^4  &=& C_1, \label{eq8}\\
\xi^1 & =& \frac{\psi r}{2}, \label{eq9}\\
\xi^1 \lambda ^\prime + 2 \xi^1 _{,1} &= &\psi,\label{eq10}
\end{eqnarray}
where $C_1$ is a constant. These consequently imply
\begin{eqnarray}
e^\nu  &=& C_2^2 r^2, \label{eq11}\\
e^\lambda & =& \left(\frac{C_3}{\psi}\right)^2, \label{eq12}\\
\xi^i& = & C_1 \delta_4^i + \left(\frac{\psi r}{2}\right)\delta_1^i,\label{eq13}
\end{eqnarray}
where $C_2$ and $C_3$ are integration constants.
Equations~(\ref{eq11})-(\ref{eq13}), help us to rewrite equations
(\ref{eq3})-(\ref{eq5}) in the form form \citep{Rahaman,Ray}
\begin{eqnarray}
\frac{1}{r^2}\left[1 - \frac{\psi^2}{C_3^2}\right]-\frac{2\psi\psi^\prime}{rC_3^2} &=& 8\pi \rho, \label{eq14}\\
\frac{1}{r^2}\left[1 - \frac{3\psi^2}{C_3^2}\right] &=& - 8\pi p_r, \label{eq15}\\
\left[\frac{\psi^2}{C_3^2r^2}\right]+\frac{2\psi\psi^\prime}{rC_3^2} &=& 8\pi p_t.\label{eq16}
\end{eqnarray}
We thus have three independent equations and four unknown parameter.
We, therefore, are free to choose any physically reasonable ansatz
for any one of these four unknown parameters to solve the set of
equations.

\subsection{Given density profile: $\rho = \frac{1}{8\pi} ( \frac{a}{r^2} + 3b )$.}
In \citet{dev1,dev2} model, an anisotropic star admits two
major types of density distributions, $\rho=$constant and
$\rho\sim r^{-2}$. These two can be constructed
in one simple form as shown above. Here $a$ and $b$ are
constants which generate various configurations of the star.
For example, by choosing $a=3/7$ and $b=0$, one may obtain a relativistic Fermi gas.
Making use of the denity profile as prescribed by  \citet{dev1,dev2},
 we rewrite equation~(\ref{eq14}) as
\begin{equation}
\frac{1}{r^2}\left[1 - \frac{\psi^2}{C_3^2}\right]-\frac{2\psi\psi^\prime}{rC_3^2}
= \left( \frac{a}{r^2} + 3b\right)\label{eq17},
\end{equation}
which can be solved easily to yield
\begin{equation}
\psi = \sqrt{\left[ (1-a)C_3^2 - C_3^2 b r^2 + \frac{C}{r}\right]}.\label{eq18}
\end{equation}
In equation (\ref{eq18}), C is an integration constant. Consequently,
 we obtain an exact analytical solution in the
form
\begin{eqnarray}
e^\nu  &=& C_2^2 r^2,\label{eq19}\\
e^\lambda & =& \frac {C_3^2} {\left[ (1-a)C_3^2 - C_3^2 b r^2
+\frac{C}{r}\right]}.\label{eq20}
\end{eqnarray}
The two pressures are obtained as
\begin{eqnarray}
p_t &=&  \frac{\left[ (1-a)C_3^2 - C_3^2 b r^2 +\frac{C}{r}\right]}{8\pi C_3^2r^2} -  \frac{\left[ bC_3^2 +\frac{C}{2r^3}\right]}{4 \pi C_3^2}, \label{eq21}\\
p_r & =&  \frac{3\left[ (1-a)C_3^2 - C_3^2 b r^2 +\frac{C}{r}\right]}{8\pi C_3^2r^2} -  \frac{1}{8 \pi r^2}.\label{eq22}
\end{eqnarray}
The measure of pressure anisotropy is given by
\begin{equation}
\Delta  =  \frac{1}{8 \pi r^2} -\frac{\left[ (1-a)C_3^2 - C_3^2 b r^2 +
\frac{C}{r}\right]}{4\pi C_3^2r^2} -  \frac{\left[ bC_3^2 +
\frac{C}{2r^3}\right]}{4 \pi C_3^2}.\label{eq23}
\end{equation}

\begin{itemize}
\item In this model, if we set $C=0$, the two metric functions $\nu(r)$ and $\lambda(r)$ become well behaved. Though the central singularity in the physical parameters like energy density, presure and anisotropic parameter can not be avoided in this formalism, the solution may be used to describe the envelope region of a star in a core-envelope type model.
\item $\frac{\Delta}{r} $ corresponds to a force due to the anisotropic
nature of the star. This force will be repulsive if $\frac{\Delta}{r} > 0$ i.e., $p_t>p_r$ and attractive if $\frac{\Delta}{r} <0$.
In the present model, $\frac{\Delta}{r}$ is given by
\begin{equation}
\frac{\Delta}{r}  =  \frac{1}{8 \pi r^3} -\frac{\left[ (1-a)C_3^2- C_3^2 b r^2 + \frac{C}{r}\right]}{4\pi C_3^2r^3} -  \frac{\left[bC_3^2 + \frac{C}{2r^3}\right]}{4 \pi C_3^2r}.\label{eq24}
\end{equation}
\item At the surface of the star $r =R$, we impose the condition that the radial pressure vanishes, i.e., $p_r(r=R)=0$, which gives
\begin{equation}
\frac{3\left[ (1-a)C_3^2 - C_3^2 b R^2 + \frac{C}{R}\right]}{8\pi C_3^2R^2} -  \frac{1}{8 \pi R^2} = 0. \label{eq25}
\end{equation}
Equation~(\ref{eq25}) can be solved easily to yield
\begin{eqnarray}
R = \left[ \frac{C}{2bC_3^2} +\sqrt{\frac{(3a-2)^3}{729b^3}+\frac{C^2}{4b^2C_2^4}}\right]^{\frac{1}{3}}+  \nonumber \\
\left[  \frac{C}{2bC_3^2} -\sqrt{\frac{(3a-2)^3}{729b^3}+\frac{C^2}{4b^2C_2^4}} \right]^{\frac{1}{3}}.\label{eq26}
\end{eqnarray}

\item If a star is composed of quark particles, then the surface tension $S$ of the star is defined as \citep{Sharma1}
$$\frac{2S}{R} = r_n (\frac{dp_r}{dr})|_{r=R},$$
where, $r_n =(\frac{1}{\pi n})^{\frac{1}{3}}$ equals the  radius of the
quark particles, $n$ is the baryon number density and $R$ is the radius of the star. By substituting the value of pressure gradient given by
\begin{equation}
8 \pi \frac{dp_r}{dr}  = \frac{2}{ r^3} -  \frac{6\left[ (1-a)C_3^2 - C_3^2 b r^2 +
\frac{C}{r}\right]}{ C_3^2r^3}- \frac{3\left[ 2bC_3^2r +
\frac{C}{r^2}\right]}{ C_3^2r^2},\label{eq27}
\end{equation}
one can calculate the surface tension in the present model. Note that $\frac{dp_r}{d\rho}$ is a decreasing
function of $r$ in this model.

\item The mass function in this case takes the form
\begin{equation}
m(r) = 4\pi \int_0^r \rho(x) x^2 dx = \frac{1}{2} ( ar
+br^3).
\end{equation}
\end{itemize}
The characteristics of the model have been shown graphically in Fig.~1-8.

\subsection{Given mass function: $\left(m(r) = \frac{br^3}{2(1+ar^2)}\right)$.}
Let us assume a mass function of the form
\begin{equation}
m(r) = \frac{br^3}{2(1+ar^2)},\label{eq28}
\end{equation}
where, $a$ and $b$ are two arbitrary constants. Such a mass function has been found to be relevant in the studies of compact stars like strange stars or dark-energy stars (see e.g., \citet{Sharma2} and references therein). As the mass $m(r)$ is defined as
\begin{equation}
m(r) = 4 \pi  \int_0^r x^2 \rho (x) dx, \label{eq29}
\end{equation}
this is equivalent to choosing the density profile in the form
\begin{equation}
8 \pi \rho = \frac{b(3+ar^2)}{(1 +ar^2)^2}.\label{30}
\end{equation}
Equation~(\ref{eq14}) for the above matter distribution takes the form
\begin{equation}
\frac{1}{r^2}\Big[1 - \frac{\psi^2}{C_3^2} \Big]-\frac{2\psi
\psi^\prime}{rC_3^2}= \frac{b(3+ar^2)}{(1 +ar^2)^2},\label{31}
\end{equation}
whose solution is given by
\begin{equation}
\psi = \sqrt{C_3^2 -  \frac {b C_3^2 r^2}{(1 + ar^2)} +\frac{C}{r}}. \label{eq32}
\end{equation}
Thus the metric functions are obtained as
\begin{eqnarray}
e^{\nu} &=& C_2 r^2, \label{eq33} \\
e^\lambda  &=& \frac {C_3^2} {\left[ C_3^2 -  \frac {b C_3^2 r^2}{(1 + ar^2)} +
\frac{C}{r}\right]}.\label{eq34}
\end{eqnarray}
Note that the metric functions are regular at the centre if we set $C=0$.

The radial and tangential pressures are obtained as
\begin{eqnarray}
8 \pi p_r  &=&  -\frac{1}{r^2} +\frac{3[C_3^2 -  \frac {b C_3^2r^2}{(1 + ar^2)} + \frac{C}{r}]}{C_3^2 r^2}, \label{eq35}\\
8 \pi p_t  &=&  \frac{[C_3^2 -  \frac {b C_3^2 r^2}{(1 + ar^2)} +\frac{C}{r}]}{r^2C_3^2 } -\frac{C} {C_3^2r^3} -\frac{2b}{(1+ar^2)^2}, \label{eq36}
\end{eqnarray}
and the measure of anisotropy is given by
\begin{equation}
8 \pi \Delta  =  \frac{1}{ r^2} -\frac{2\left[ C_3^2 -  \frac {b C_3^2 r^2}{(1 + ar^2)} +
\frac{C}{r}\right]}{ C_3^2r^2} -\frac{C} {C_3^2r^3} -\frac{2b}
{(1+ar^2)^2}.\label{eq37}
\end{equation}
At the boundary, radial pressure vanishes ($p_r(r=R)=0$), which gives
\begin{equation}
 -\frac{1}{R^2} +\frac{3[C_3^2 -  \frac {b C_3^2
R^2}{(1 + aR^2)} + \frac{C}{R}]}{C_3^2 R^2} = 0. \label{eq38}
\end{equation}
Equation~(\ref{eq38}) determines the radius $R$ of the star.

We, thus, obtain all the physical parameters in simple analytic forms.
 Though energy density is regular throughout the interior of a star,
the two pressures still remain singular in this model. The
characteristics of the model are shown graphically in Fig.~9-16.

\section{Discussions}
It is an established fact that the density within a compact star may go beyond nuclear density and anisotropy may develope at the interior of the star. To model a compact star with highly anisotropic matter distribution, we require a relativistic treatment. In this model, the anisotropic star is assumed to be a spherically symmetric fluid distribution where the total
pressure of the fluid is decomposed into two pressure terms, the radial and
the transverse one. The difference between the two is a measure of the surface tension of the star, vis-a-vis stiffness of the core.

Here we have obtained a new class of solutions in simple analytical
forms describing anisotropic stars admitting conformal motion. These
are obtained by taking energy density and mass variation profiles in
two different cases. The solutions obtained here are in simple
closed forms and can be used to study the physical behavior of
compact anisotropic stars like neutron stars and strange stars. For
a physically meaningful solution, it is imperative to study the
behaviour of the physical parameters like energy density, mass,
pressure gradient and force inside the star. In this model all these
parameters are well-behaved as shown in Fig.~1-16. However, our
solutions suffer from a central singularity problem as can be found
in some earlier works \citep{mak,Aktas,Rahaman} as well and,
therefore, is suitable for the description of the envelope region of
the star. It will be interesting to examine whether other forms of
matter distributions can solve the central singularity problem for
anisotropic stars possessing conformal symmetry. This will be taken
up elsewhere.

\acknowledgments
RS gratefully acknowledges support from IUCAA, Pune, India, under the Visiting Research Associateship Programme.

\begin{figure}[t]
\includegraphics[scale=.33]{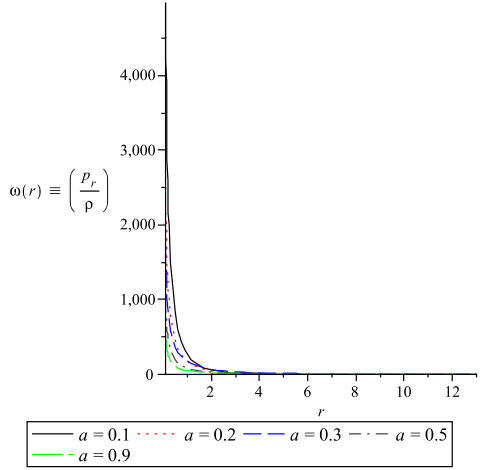}\\
\caption{The EoS parameter $\omega$ is plotted against the radial parameter.
 Chosen parameters are $C=.5$, $C_3 = .1$ and $b =.03$. Radius of the star is $R=13$ km.}
\end{figure}
\begin{figure} \centering
\includegraphics[scale=.33]{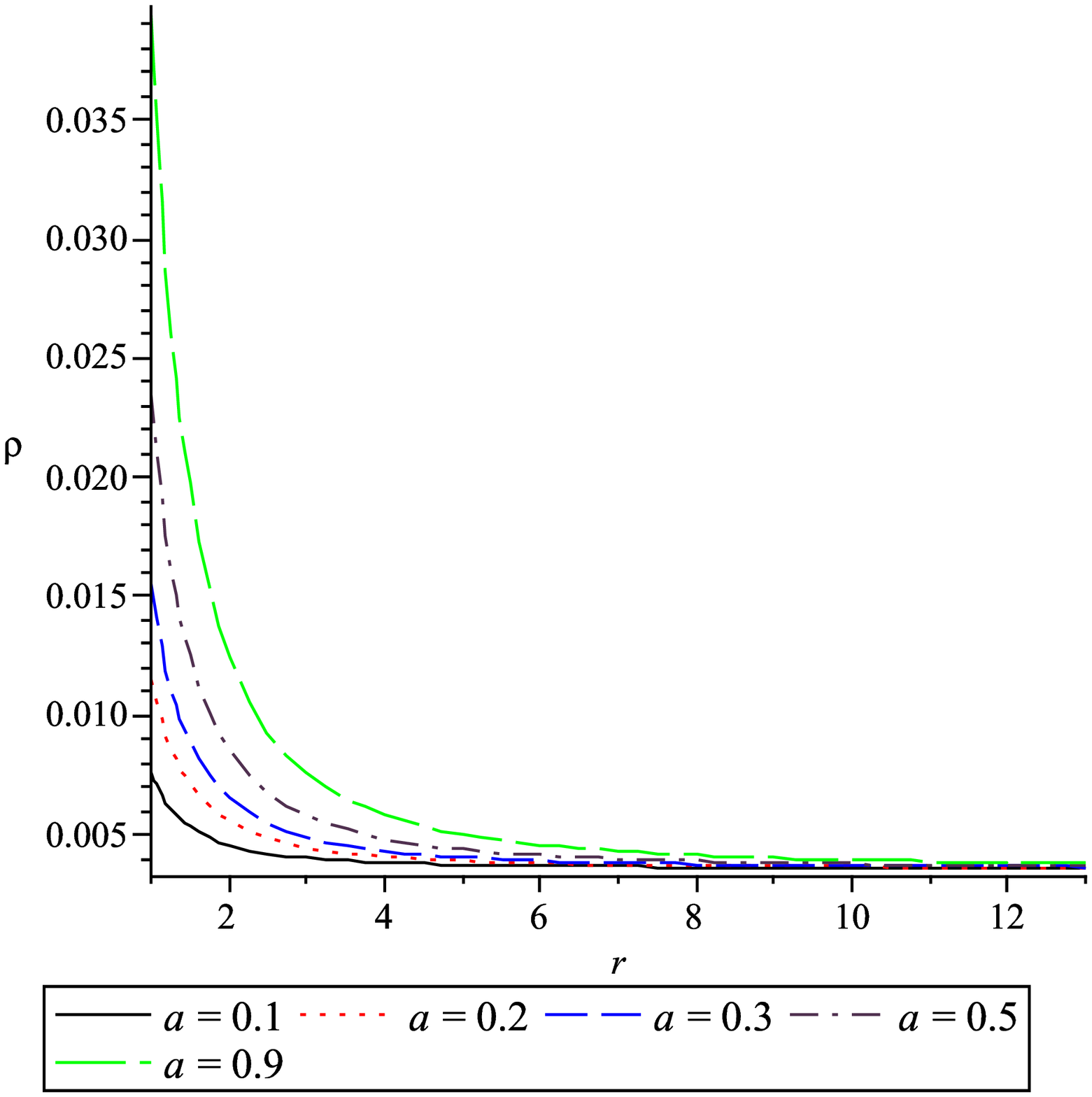}\\
\caption{The density parameter $\rho$ is shown against $r$. Chosen
parameters are the same as in Fig.1}
\end{figure}\begin{figure} \centering
\includegraphics[scale=.33]{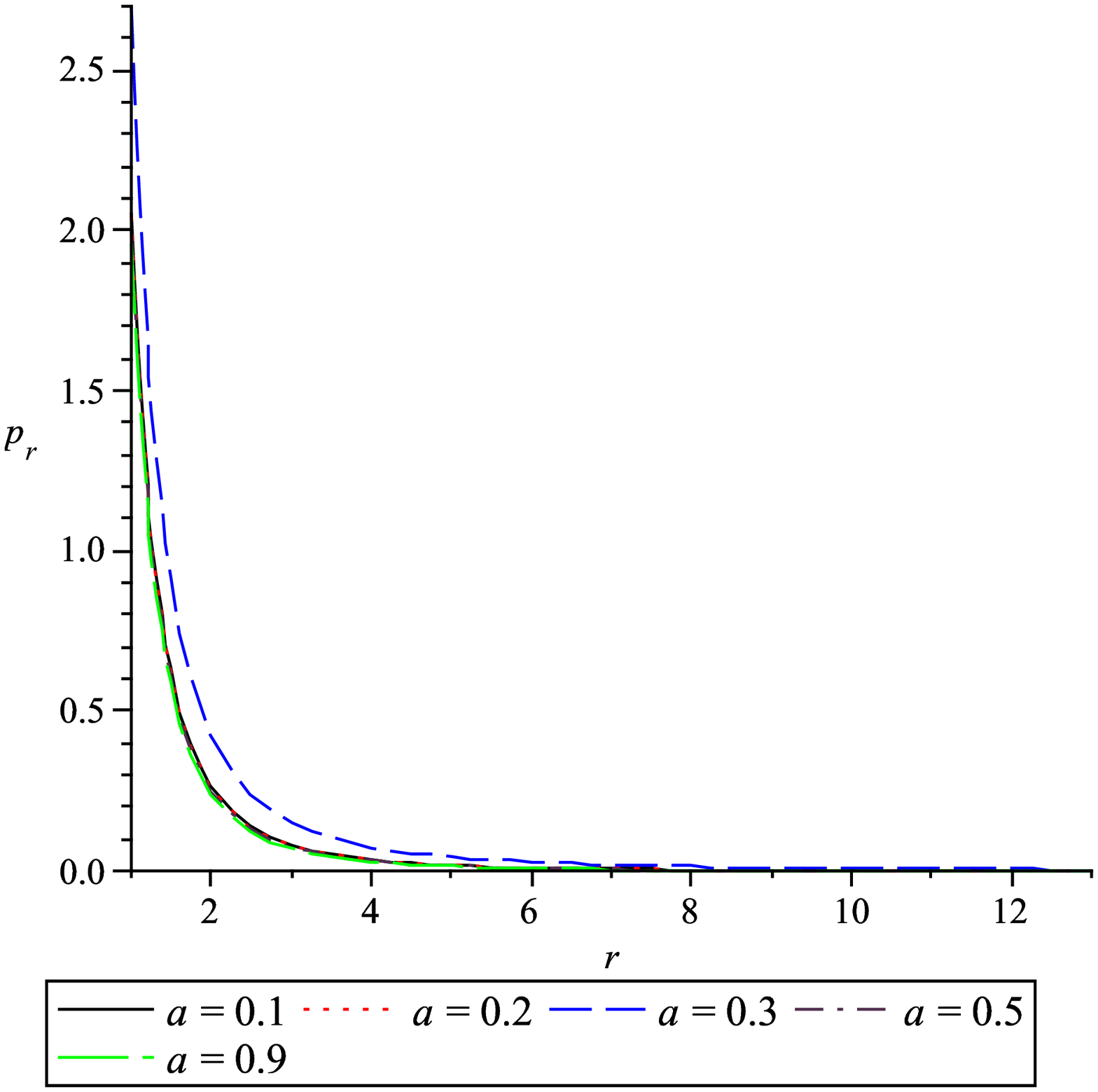}\\
\caption{The radial pressure $p_r$ is shown against $r$. Chosen
parameters are the same as in Fig.1}
\end{figure}\begin{figure} \centering
\includegraphics[scale=.33]{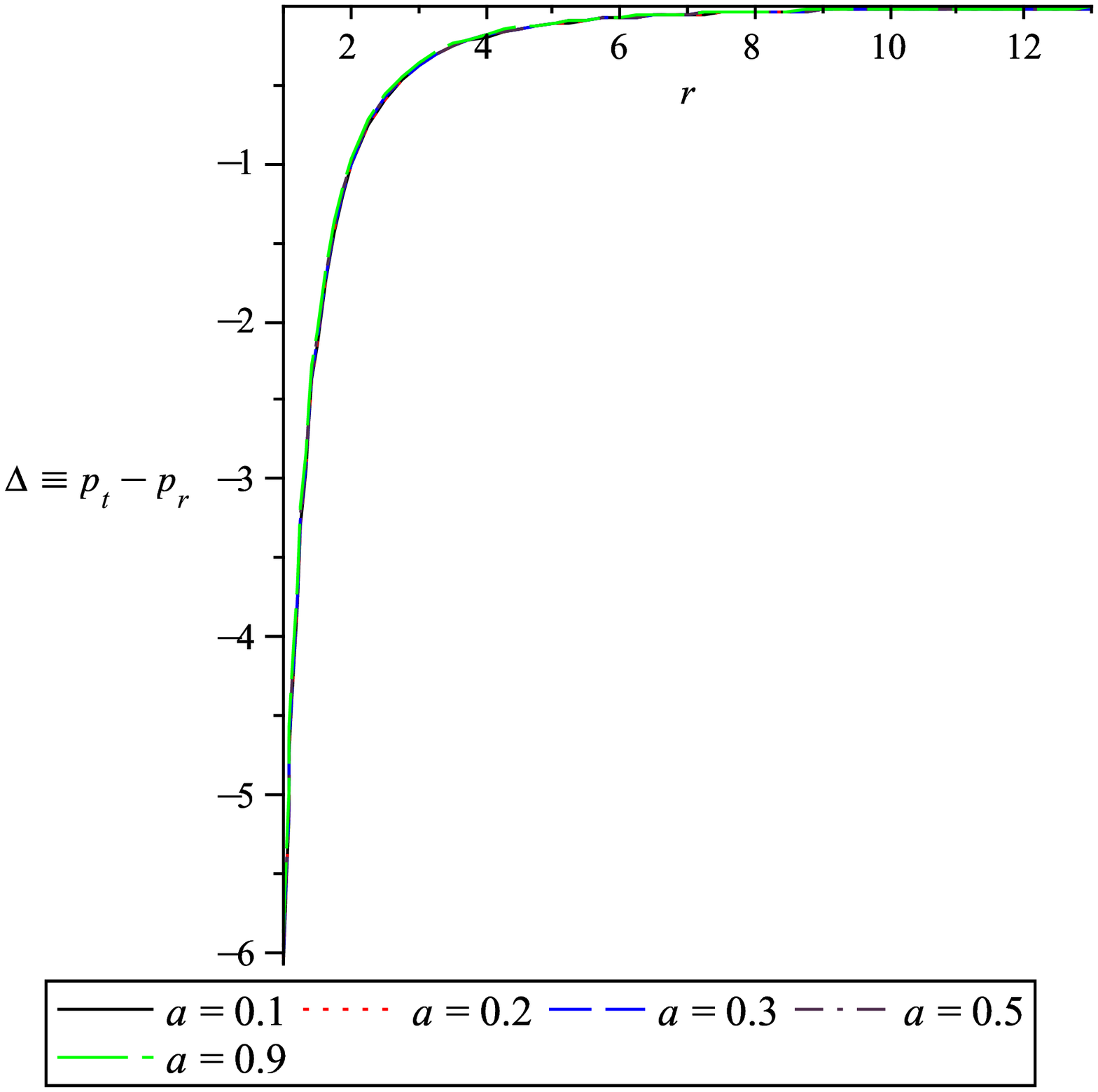}\\
\caption{The anisotropy $p_t-p_r$ is shown against $r$. Chosen
parameters are the same as in Fig.1}
\end{figure}\begin{figure} \centering
\includegraphics[scale=.33]{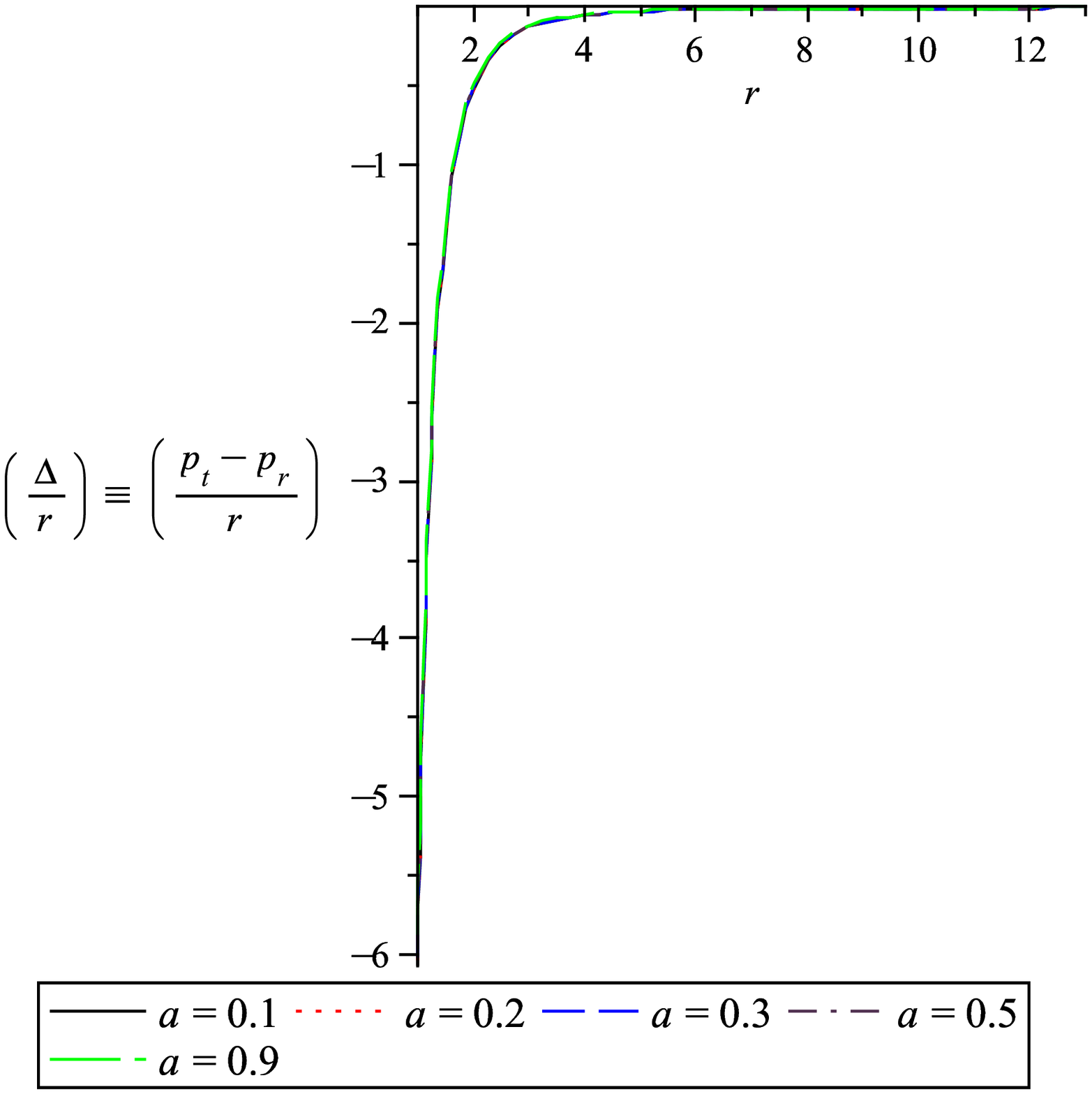}\\
\caption{The force parameter is shown against $r$. Chosen parameters
are the same as in Fig.1}
\end{figure}\begin{figure} \centering
\includegraphics[scale=.33]{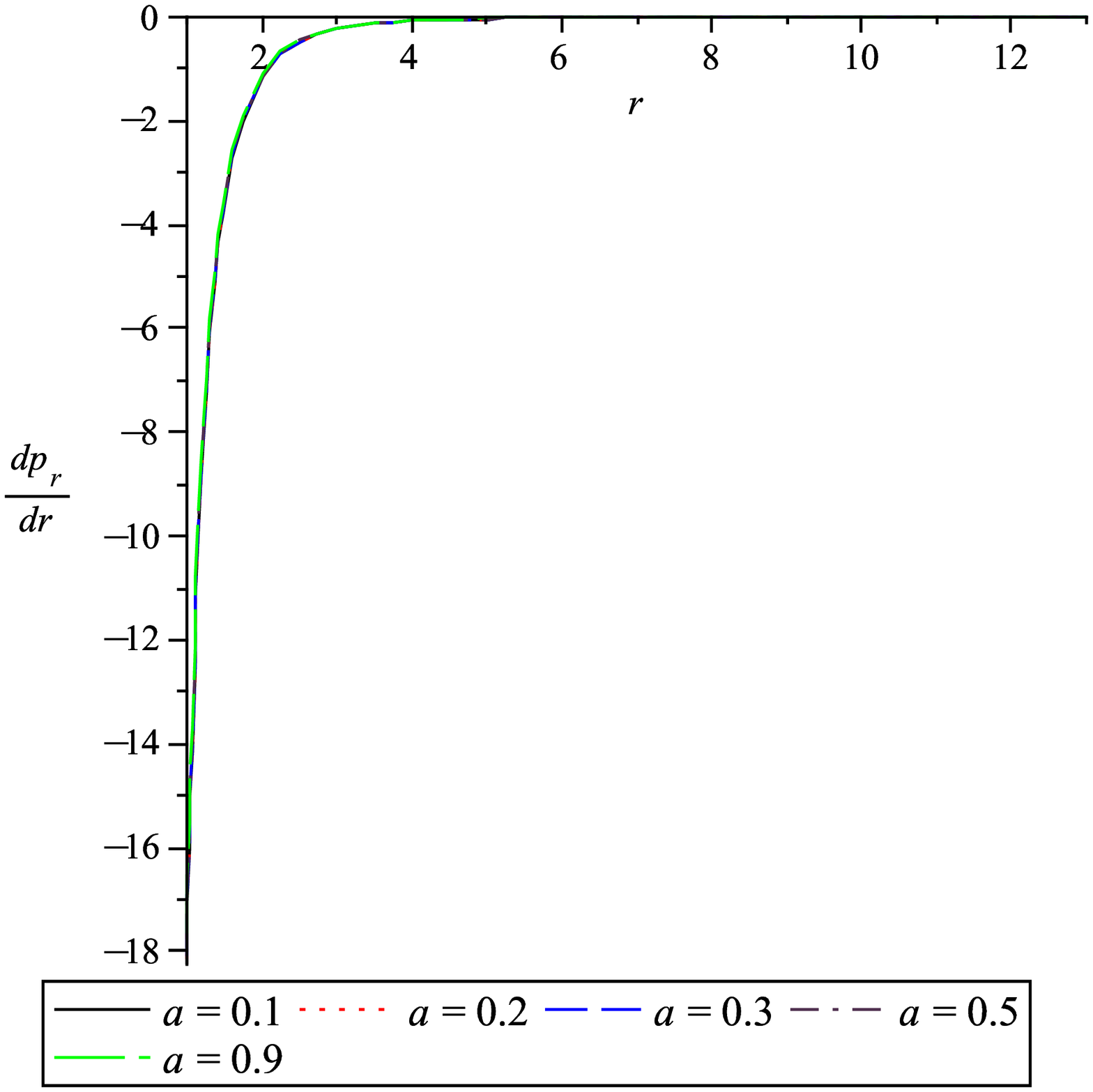}\\
\caption{The radial pressure gradient is shown against $r$.
 Chosen parameters are the same as in Fig.1}
\end{figure}\begin{figure}\centering
\includegraphics[scale=.33]{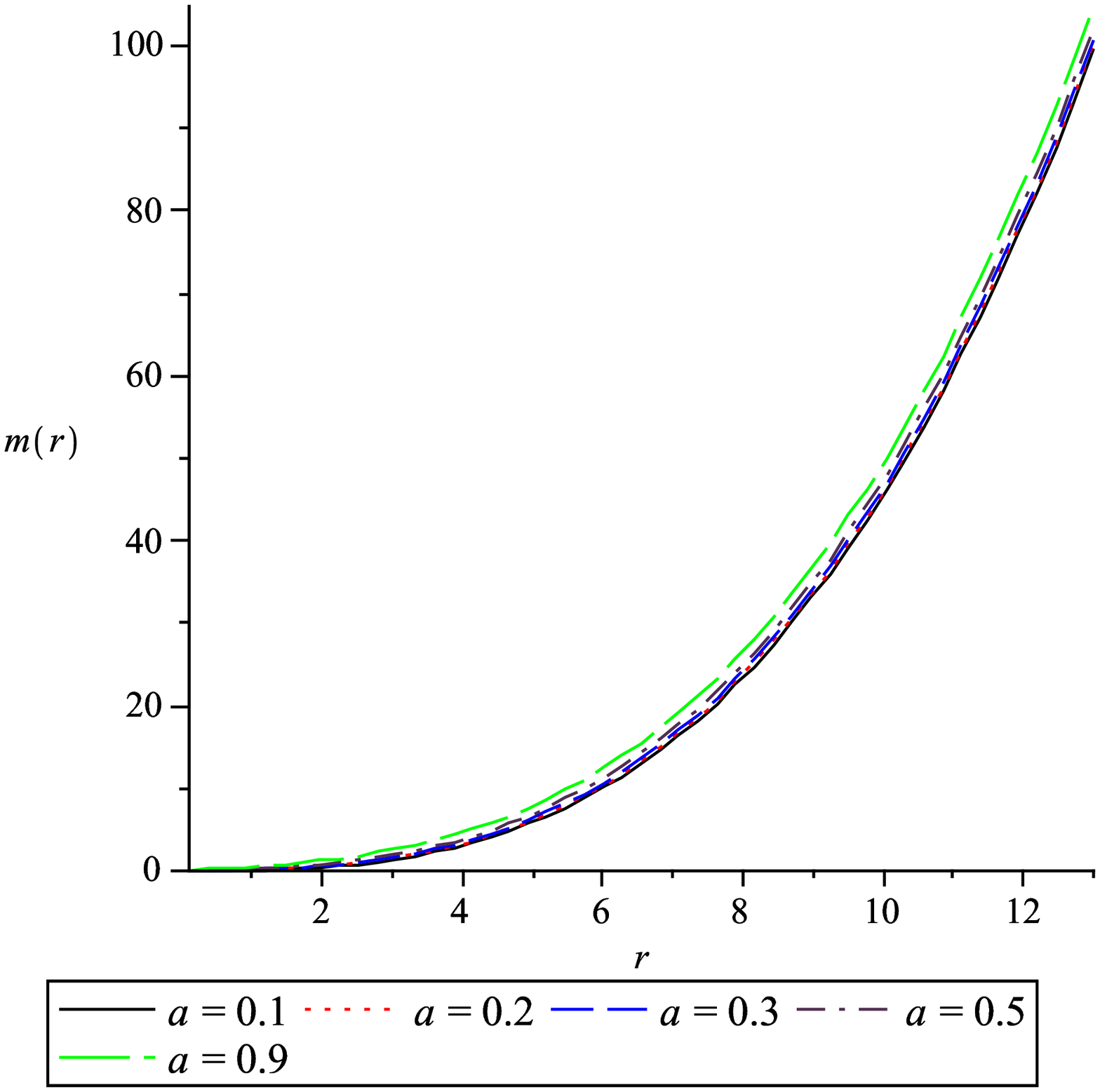}\\
\caption{The mass parameter $m(r)$ is shown against $r$. Chosen
parameters are the same as in Fig.1}
\end{figure}\begin{figure}\centering
\includegraphics[scale=.33]{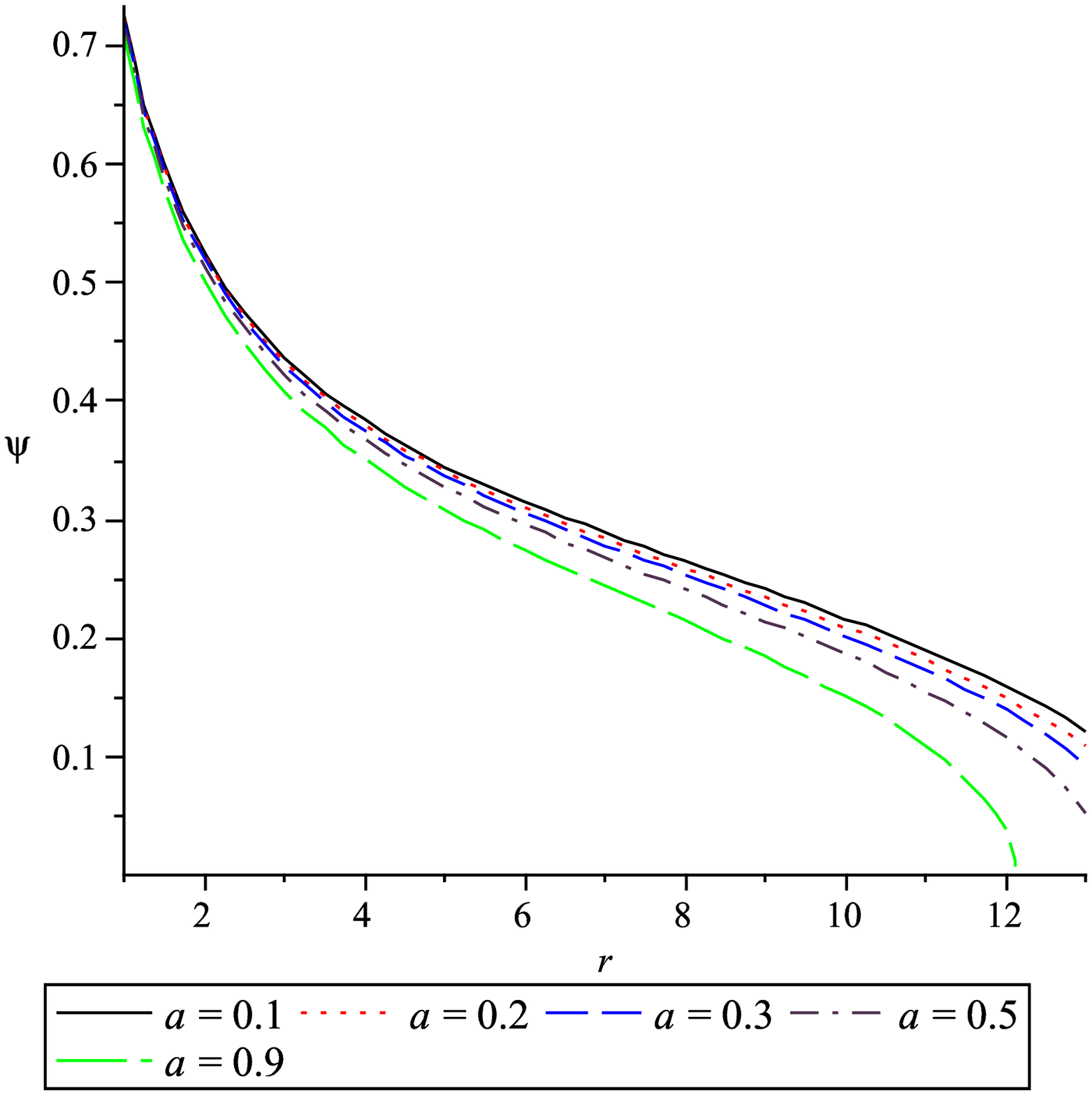}\\
\caption{The conformal parameter $\psi(r)$ is shown against $r$.
Chosen parameters are the same as in Fig.1}
\end{figure}
\begin{figure} \centering
\includegraphics[scale=.33]{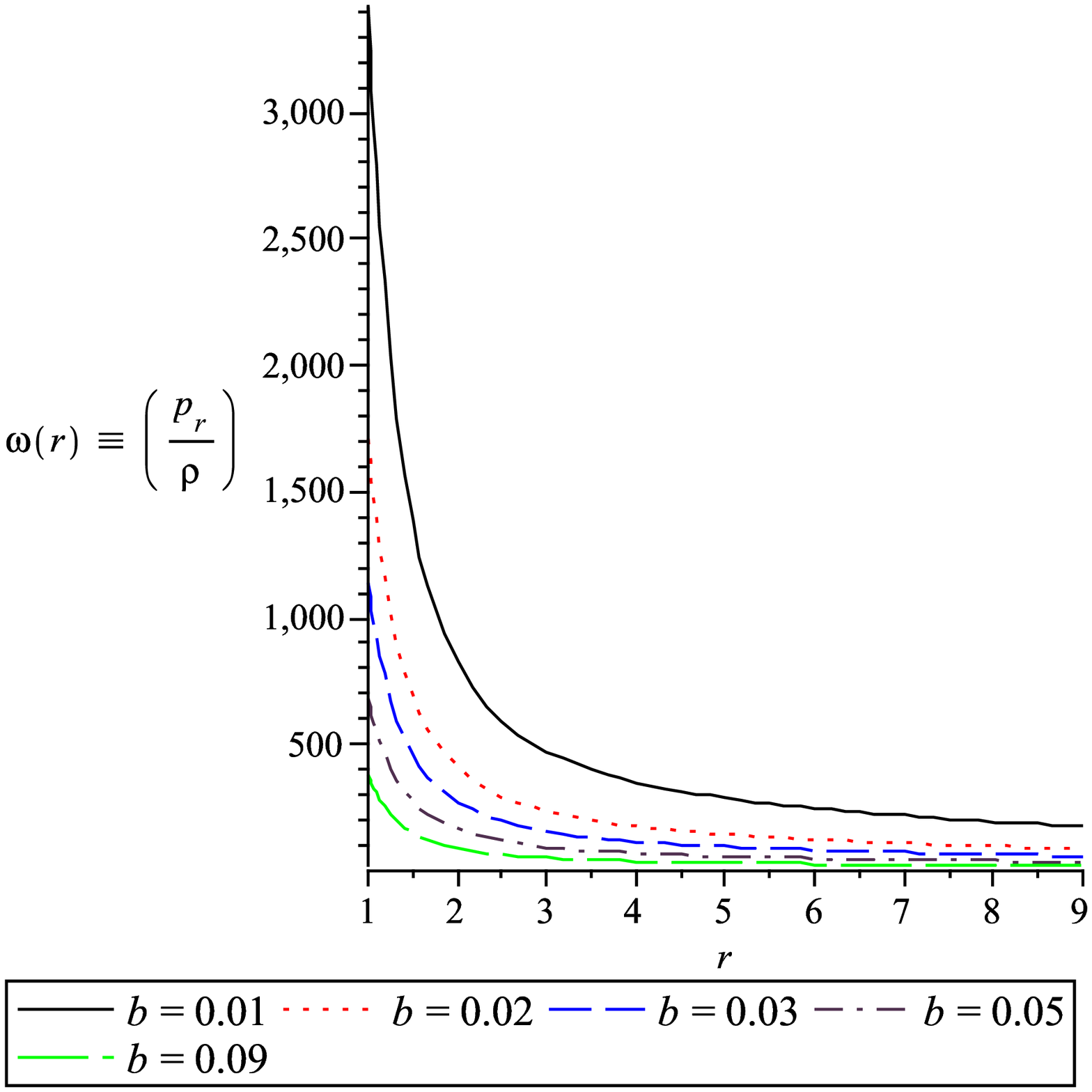}\\
\caption{\label{fg1}The EoS parameter $\omega$ is plotted against
the radial parameter.
 Chosen parameters are $C=.5$, $C_3 = .1$ and $a =.2$. Radius of the star is $R=9$ km.}
\end{figure}\begin{figure} \centering
\includegraphics[scale=.33]{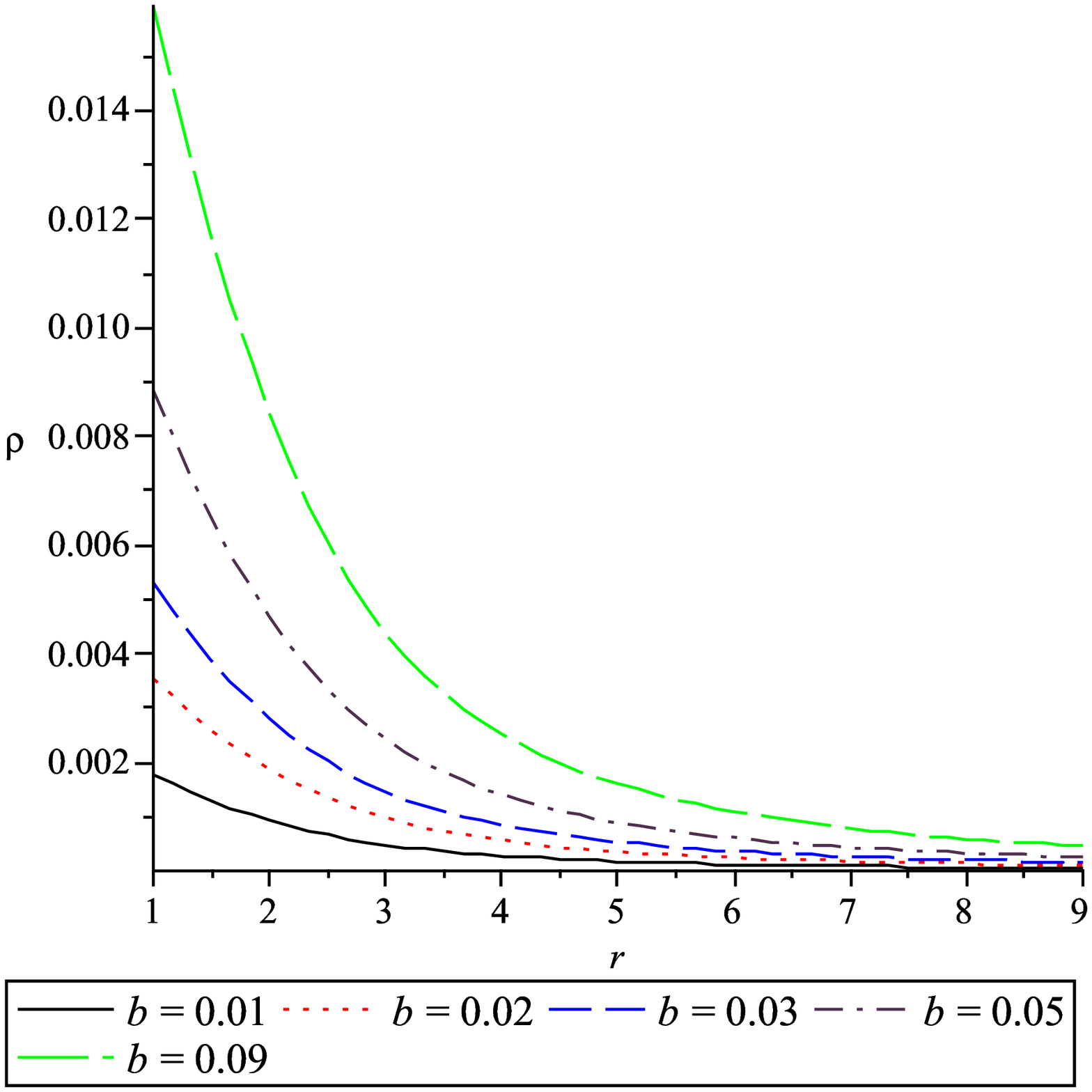}\\
\caption{The density parameter $\rho$ is shown against $r$. Chosen
parameters are the same as in Fig. 9}
\end{figure}\begin{figure} \centering
\includegraphics[scale=.33]{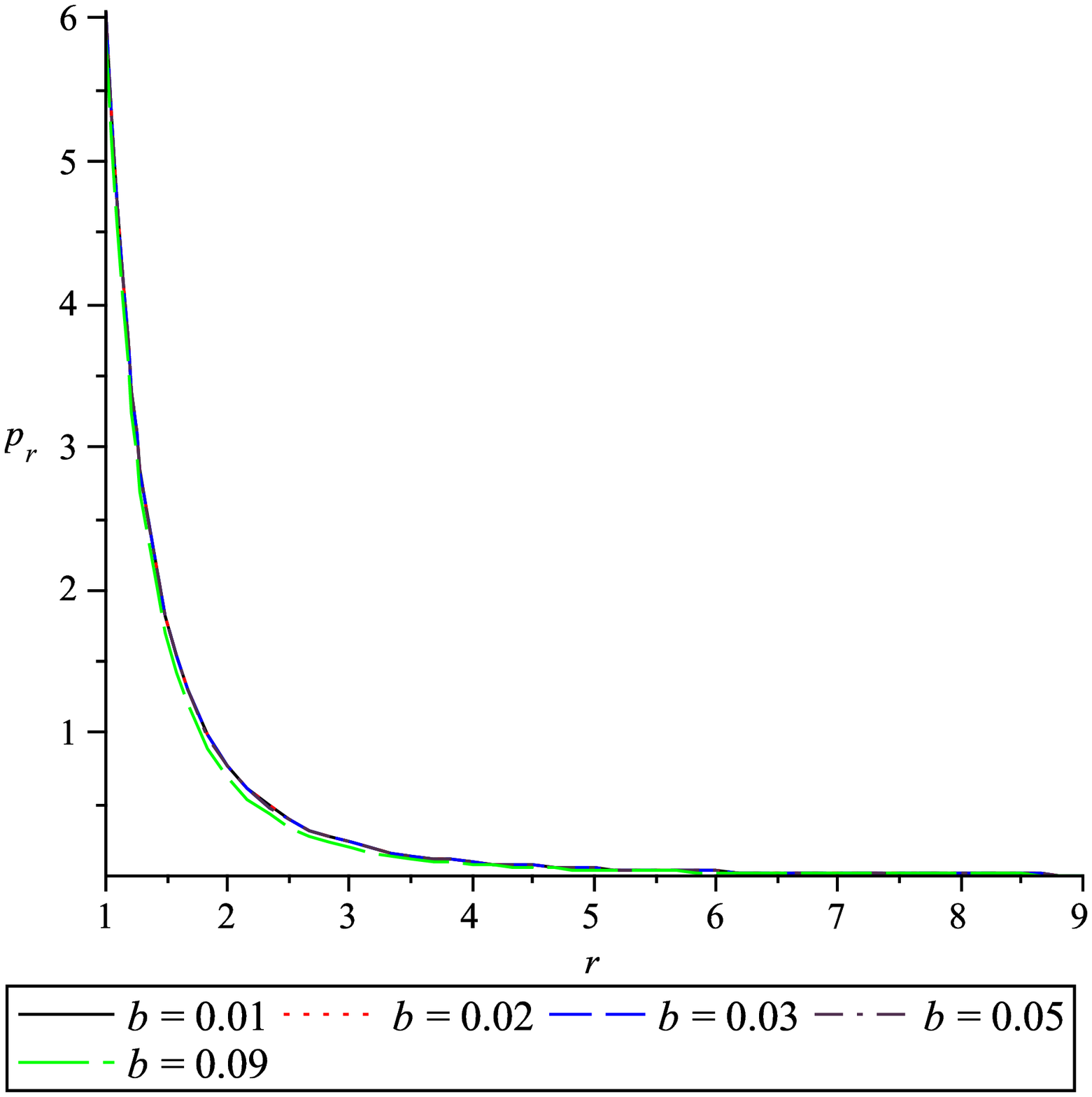}\\
\caption{The radial pressure $p_r$ is shown against $r$. Chosen
parameters are the same as in Fig. 9}
\end{figure}\begin{figure} \centering
\includegraphics[scale=.33]{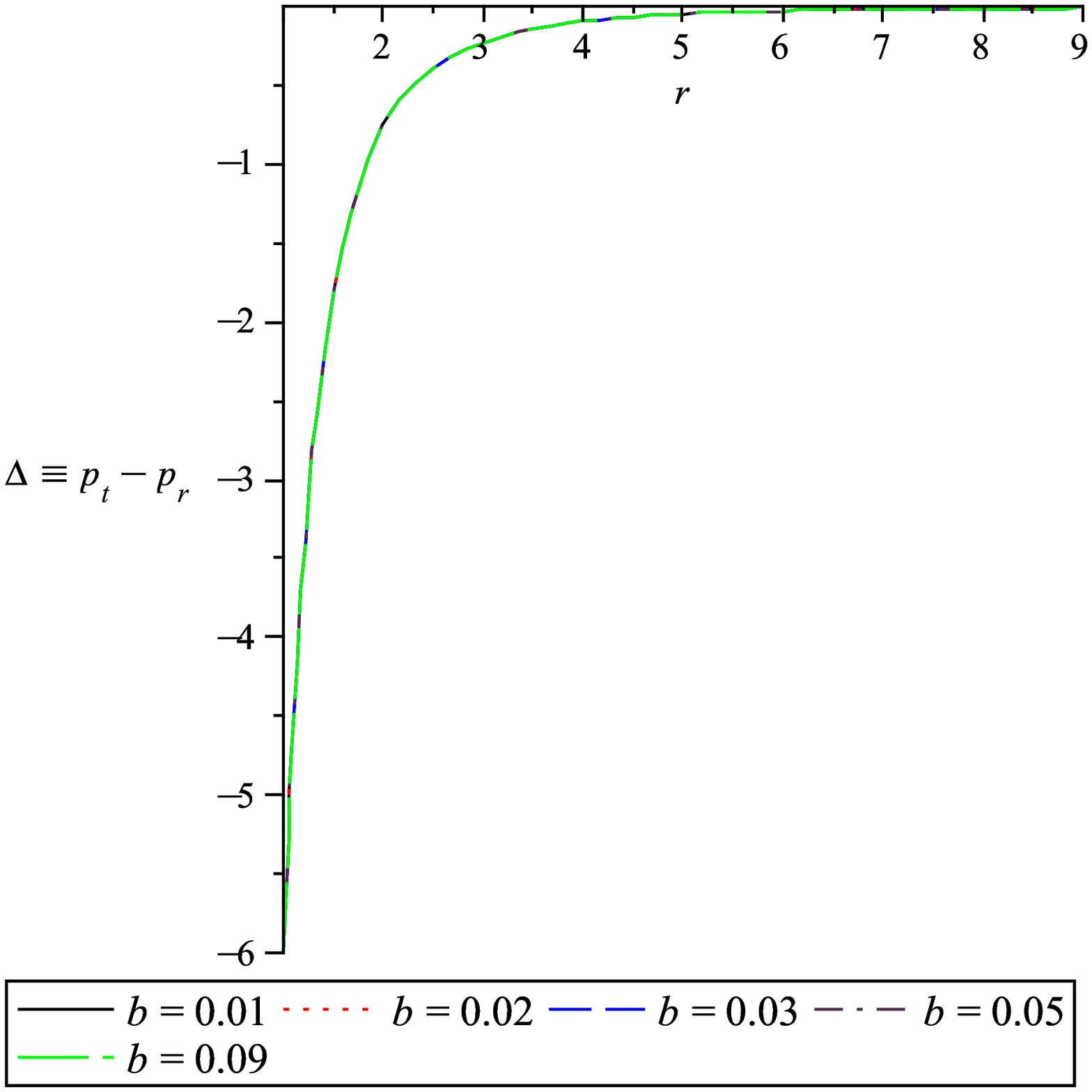}\\
\caption{The anisotropy $p_t-p_r$ is shown against $r$. Chosen
parameters are the same as in Fig. 9}
\end{figure}\begin{figure} \centering
\includegraphics[scale=.33]{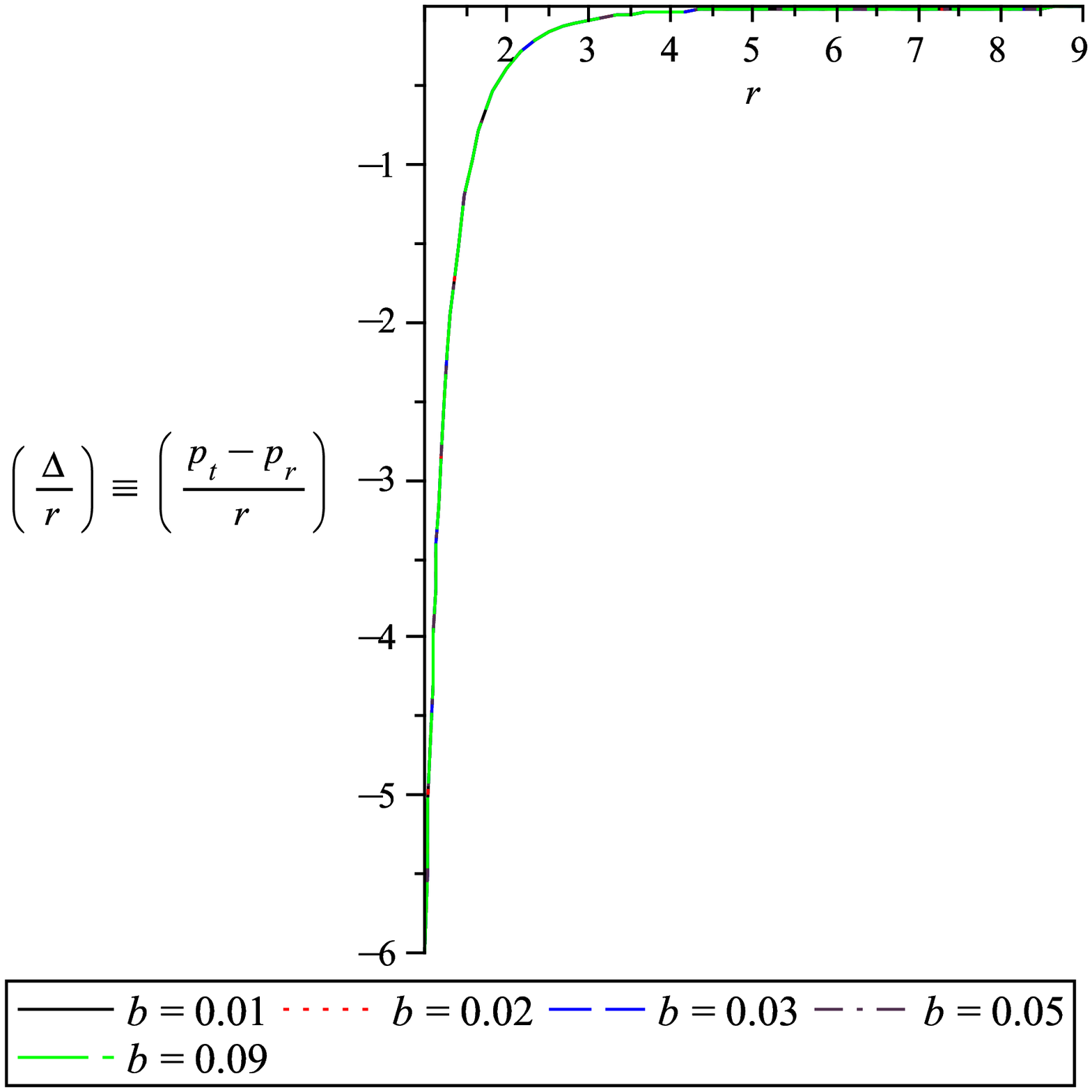}\\
\caption{The force parameter is shown against $r$. Chosen parameters
are the same as in Fig. 9}
\end{figure}\begin{figure} \centering
\includegraphics[scale=.33]{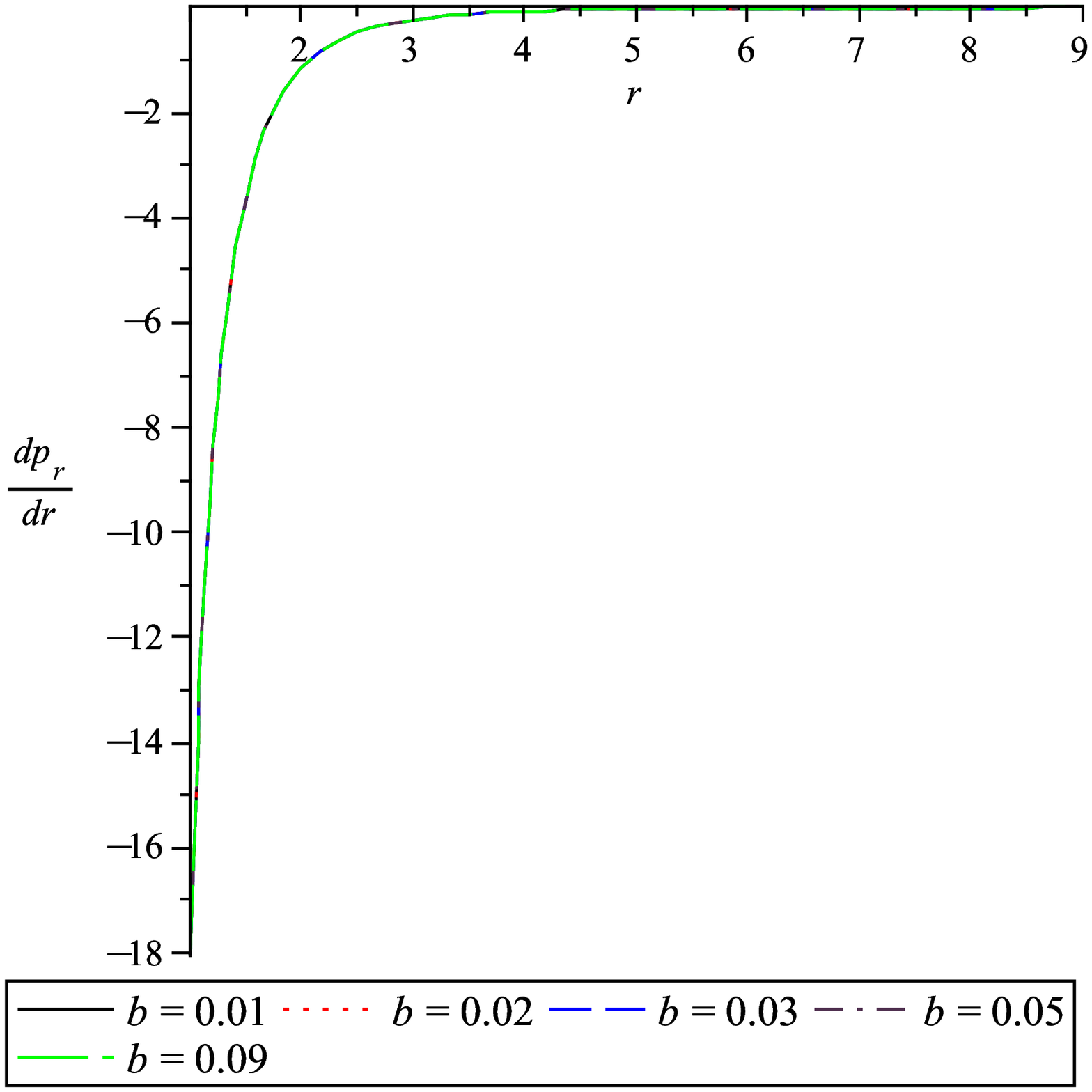}\\
\caption{The radial pressure gradient is shown against $r$.
 Chosen parameters are the same as in Fig. 9}
\end{figure}\begin{figure}\centering
\includegraphics[scale=.33]{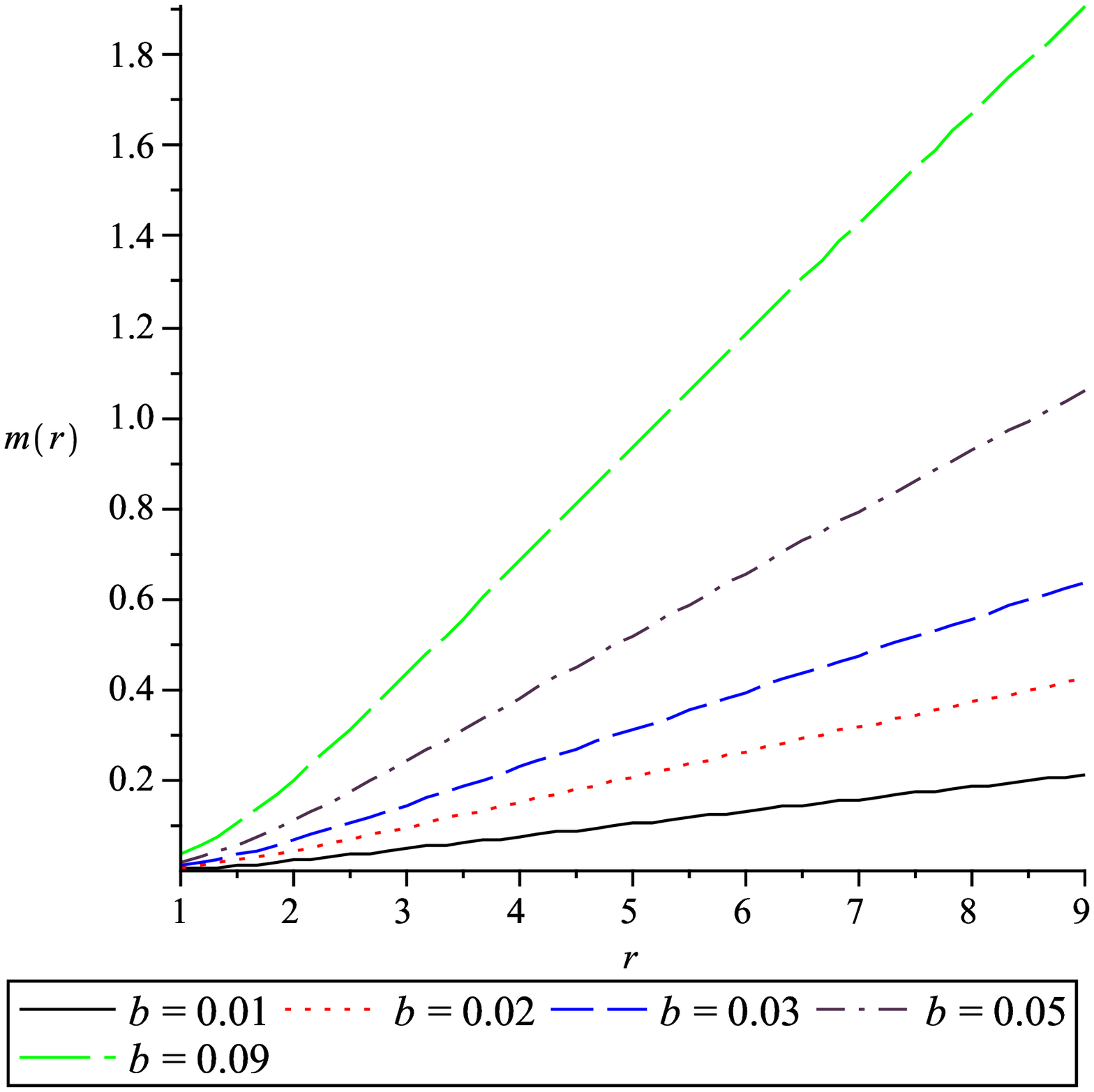}\\
\caption{The mass parameter $m(r)$ is shown against $r$. Chosen
parameters are the same as in Fig. 9}
\end{figure}\begin{figure}\centering
\includegraphics[scale=.33]{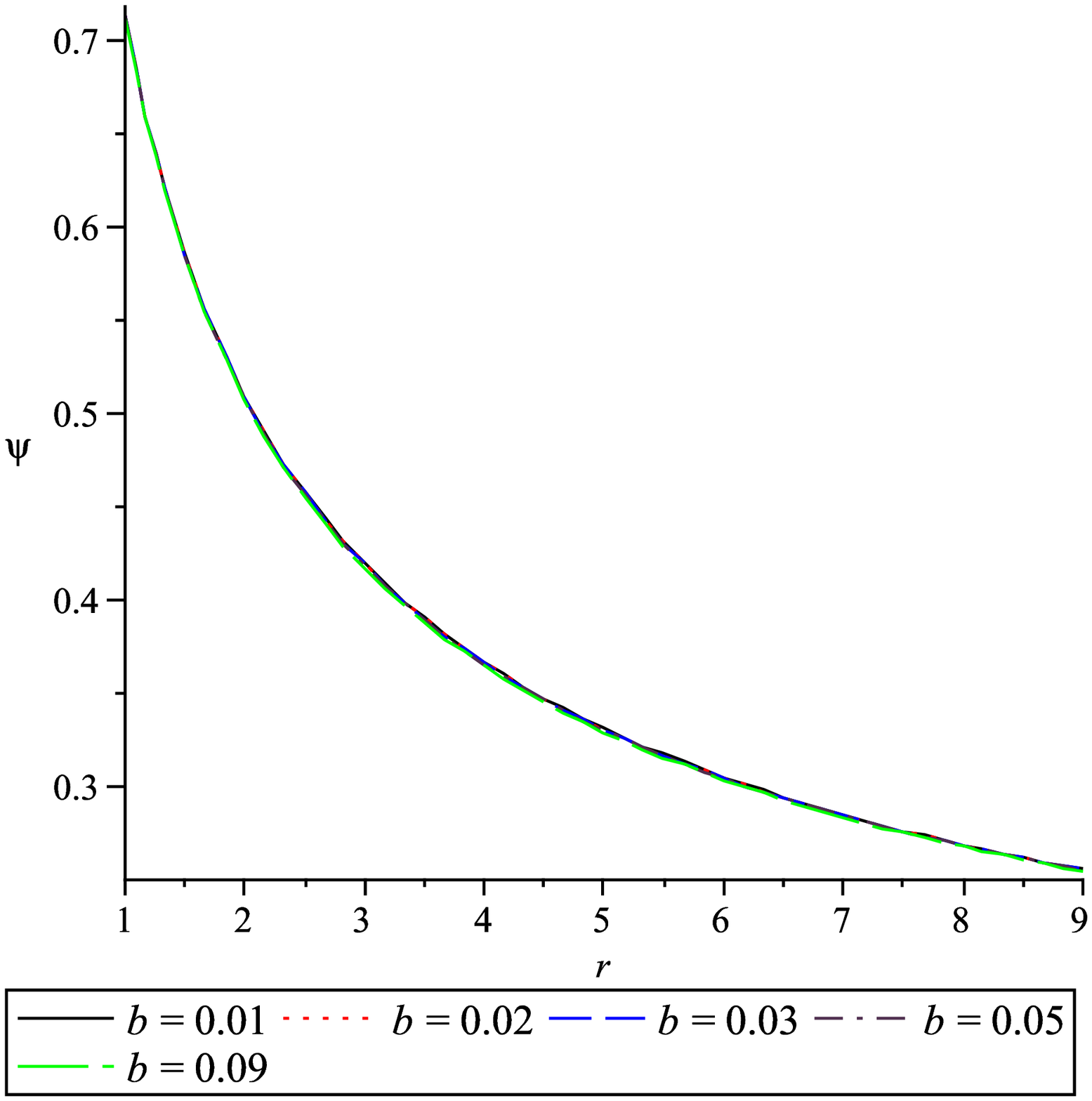}\\
\caption{The conformal parameter $\psi(r)$ is shown against $r$.
Chosen parameters are the same as in Fig. 9}
\end{figure}

\end{document}